\documentstyle[12pt]{article}

\catcode`\@=11
\long\def\@makefntext#1{ 
\protect\noindent \hbox to 3.2pt {\hskip-.9pt
$^{{\ninerm\@thefnmark}}$\hfil}#1\hfill} 

\def\thefootnote{\fnsymbol{footnote}}
 \def\@makefnmark{\hbox to 0pt{$^{\@thefnmark}$\hss}}  

\def\ps@myheadings{\let\@mkboth\@gobbletwo
\def\@oddhead{\hbox{} 
\rightmark\hfil\ninerm\thepage}
\def\@oddfoot{}\def\@evenhead{\ninerm\thepage\hfil 
\leftmark\hbox{}}\def\@evenfoot{}
\def\sectionmark##1{}\def\subsectionmark##1{}}

\begin{document}

\newcommand{\symbolfootnote}{\renewcommand{\thefootnote}
        {\fnsymbol{footnote}}}
\renewcommand{\thefootnote}{\fnsymbol{footnote}}
\newcommand{\alphfootnote}
        {\setcounter{footnote}{0}
         \renewcommand{\thefootnote}{\sevenrm\alph{footnote}}}

\newcounter{sectionc}\newcounter{subsectionc}\newcounter{subsubsectionc}
\renewcommand{\section}[1] {\vspace{0.6cm}\addtocounter{sectionc}{1}
\setcounter{subsectionc}{0}\setcounter{subsubsectionc}{0}\noindent
        {\bf\thesectionc. #1}\par\vspace{0.4cm}}
\renewcommand{\subsection}[1] {\vspace{0.6cm}\addtocounter{subsectionc}{1}
        \setcounter{subsubsectionc}{0}\noindent
        {\it\thesectionc.\thesubsectionc. #1}\par\vspace{0.4cm}}
\renewcommand{\subsubsection}[1]
{\vspace{0.6cm}\addtocounter{subsubsectionc}{1}
        \noindent {\rm\thesectionc.\thesubsectionc.\thesubsubsectionc.
        #1}\par\vspace{0.4cm}}
\newcommand{\nonumsection}[1] {\vspace{0.6cm}\noindent{\bf #1}
        \par\vspace{0.4cm}}

\newcounter{appendixc}
\newcounter{subappendixc}[appendixc]
\newcounter{subsubappendixc}[subappendixc]
\renewcommand{\thesubappendixc}{\Alph{appendixc}.\arabic{subappendixc}}
\renewcommand{\thesubsubappendixc}
        {\Alph{appendixc}.\arabic{subappendixc}.\arabic{subsubappendixc}}

\renewcommand{\appendix}[1] {\vspace{0.6cm}
        \refstepcounter{appendixc}
        \setcounter{figure}{0}
        \setcounter{table}{0}
        \setcounter{equation}{0}
        \renewcommand{\thefigure}{\Alph{appendixc}.\arabic{figure}}
        \renewcommand{\thetable}{\Alph{appendixc}.\arabic{table}}
        \renewcommand{\theappendixc}{\Alph{appendixc}}
        \renewcommand{\theequation}{\Alph{appendixc}.\arabic{equation}}
        \noindent{\bf Appendix \theappendixc #1}\par\vspace{0.4cm}}
\newcommand{\subappendix}[1] {\vspace{0.6cm}
        \refstepcounter{subappendixc}
        \noindent{\bf Appendix \thesubappendixc. #1}\par\vspace{0.4cm}}
\newcommand{\subsubappendix}[1] {\vspace{0.6cm}
        \refstepcounter{subsubappendixc}
        \noindent{\it Appendix \thesubsubappendixc. #1}
        \par\vspace{0.4cm}}

\def\abstracts#1{{
        \centering{\begin{minipage}{30pc}\tenrm\baselineskip=12pt\noindent
        \centerline{\tenrm ABSTRACT}\vspace{0.3cm}
        \parindent=0pt #1
        \end{minipage} }\par}}

\newcommand{\bibit}{\it}
\newcommand{\bibbf}{\bf}
\renewenvironment{thebibliography}[1]
        {\begin{list}{\arabic{enumi}.}
        {\usecounter{enumi}\setlength{\parsep}{0pt}
\setlength{\leftmargin 1.25cm}{\rightmargin 0pt}
         \setlength{\itemsep}{0pt} \settowidth
        {\labelwidth}{#1.}\sloppy}}{\end{list}}

\topsep=0in\parsep=0in\itemsep=0in
\parindent=1.5pc

\newcounter{itemlistc}
\newcounter{romanlistc}
\newcounter{alphlistc}
\newcounter{arabiclistc}
\newenvironment{itemlist}
        {\setcounter{itemlistc}{0}
         \begin{list}{$\bullet$}
        {\usecounter{itemlistc}
         \setlength{\parsep}{0pt}
         \setlength{\itemsep}{0pt}}}{\end{list}}

\newenvironment{romanlist}
        {\setcounter{romanlistc}{0}
         \begin{list}{$($\roman{romanlistc}$)$}
        {\usecounter{romanlistc}
         \setlength{\parsep}{0pt}
         \setlength{\itemsep}{0pt}}}{\end{list}}

\newenvironment{alphlist}
        {\setcounter{alphlistc}{0}
         \begin{list}{$($\alph{alphlistc}$)$}
        {\usecounter{alphlistc}
         \setlength{\parsep}{0pt}
         \setlength{\itemsep}{0pt}}}{\end{list}}

\newenvironment{arabiclist}
        {\setcounter{arabiclistc}{0}
         \begin{list}{\arabic{arabiclistc}}
        {\usecounter{arabiclistc}
         \setlength{\parsep}{0pt}
         \setlength{\itemsep}{0pt}}}{\end{list}}

\newcommand{\fcaption}[1]{
        \refstepcounter{figure}
        \setbox\@tempboxa = \hbox{\tenrm Fig.~\thefigure. #1}
        \ifdim \wd\@tempboxa > 6in
           {\begin{center}
        \parbox{6in}{\tenrm\baselineskip=12pt Fig.~\thefigure. #1 }
            \end{center}}
        \else
             {\begin{center}
             {\tenrm Fig.~\thefigure. #1}
              \end{center}}
        \fi}

\newcommand{\tcaption}[1]{
        \refstepcounter{table}
        \setbox\@tempboxa = \hbox{\tenrm Table~\thetable. #1}
        \ifdim \wd\@tempboxa > 6in
           {\begin{center}
        \parbox{6in}{\tenrm\baselineskip=12pt Table~\thetable. #1 }
            \end{center}}
        \else
             {\begin{center}
             {\tenrm Table~\thetable. #1}
              \end{center}}
        \fi}

\def\@citex[#1]#2{\if@filesw\immediate\write\@auxout
        {\string\citation{#2}}\fi
\def\@citea{}\@cite{\@for\@citeb:=#2\do
        {\@citea\def\@citea{,}\@ifundefined
        {b@\@citeb}{{\bf ?}\@warning
        {Citation `\@citeb' on page \thepage \space undefined}}
        {\csname b@\@citeb\endcsname}}}{#1}}

\newif\if@cghi
\def\cite{\@cghitrue\@ifnextchar [{\@tempswatrue
        \@citex}{\@tempswafalse\@citex[]}}
\def\citelow{\@cghifalse\@ifnextchar [{\@tempswatrue
        \@citex}{\@tempswafalse\@citex[]}}
\def\@cite#1#2{{$\null^{#1}$\if@tempswa\typeout
        {IJCGA warning: optional citation argument
        ignored: `#2'} \fi}}
\newcommand{\citeup}{\cite}

\def\fnm#1{$^{\mbox{\scriptsize #1}}$}
\def\fnt#1#2{\footnotetext{\kern-.3em
        {$^{\mbox{\sevenrm #1}}$}{#2}}}

\font\twelvebf=cmbx10 scaled\magstep 1
\font\twelverm=cmr10 scaled\magstep 1
\font\twelveit=cmti10 scaled\magstep 1
\font\elevenbfit=cmbxti10 scaled\magstephalf
\font\elevenbf=cmbx10 scaled\magstephalf
\font\elevenrm=cmr10 scaled\magstephalf
\font\elevenit=cmti10 scaled\magstephalf
\font\bfit=cmbxti10
\font\tenbf=cmbx10
\font\tenrm=cmr10
\font\tenit=cmti10
\font\ninebf=cmbx9
\font\ninerm=cmr9
\font\nineit=cmti9
\font\eightbf=cmbx8
\font\eightrm=cmr8
\font\eightit=cmti8


\centerline{\tenbf QUANTUM ALGEBRAIC SYMMETRIES IN NUCLEAR STRUCTURE}
\vspace{0.8cm}
\centerline{\tenrm Dennis BONATSOS}
\baselineskip=13pt
\centerline{\tenit ECT$^*$, Villa Tambosi, Strada delle Tabarelle 286}
\baselineskip=12pt
\centerline{\tenit I-38050 Villazzano (Trento), Italy}
\vspace{0.3cm}
\centerline{\tenrm C. DASKALOYANNIS}
\baselineskip=13pt
\centerline{\tenit Department of Physics, Aristotle University of Thessaloniki}
\baselineskip=12pt
\centerline{\tenit GR-54006 Thessaloniki, Greece}
\vspace{0.3cm}
\centerline{\tenrm P. KOLOKOTRONIS, D. LENIS}
\baselineskip=13pt
\centerline{\tenit Institute of Nuclear Physics, NCSR ``Demokritos''}
\baselineskip=12pt
\centerline{\tenit GR-15310 Aghia Paraskevi, Attiki, Greece}
\vspace{0.9cm}
\abstracts{Various applications of quantum algebraic techniques in nuclear
structure physics, such as the su$_q$(2) rotator model and its extensions,
the use of deformed bosons in the description of pairing correlations, and
the construction of deformed exactly soluble models (Interacting Boson Model,
Moszkowski model) are briefly reviewed. Emphasis is put in the study of
the symmetries of the anisotropic quantum harmonic oscillator with rational
ratios of frequencies, which underlie the structure of superdeformed and
hyperdeformed nuclei, the Bloch--Brink $\alpha$-cluster model and possibly
the shell structure in deformed atomic clusters.
}

\vfil
\rm\baselineskip=14pt
\section{Introduction}

Quantum algebras $^{1,2}$ (also called quantum groups) are deformed versions
of the
usual Lie algebras, to which they reduce when the deformation parameter
$q$ is set equal to unity.
{}From the mathematical point of view they have the structure of Hopf algebras
$^3$.
Their use in physics became popular with the
introduction $^{4-6}$ of the $q$-deformed harmonic oscillator as a tool for
providing a boson realization of the quantum algebra su$_q$(2), although
similar mathematical structures had already been known $^{7,8}$.
Initially used for solving the quantum Yang--Baxter equation, quantum algebras
have subsequently found applications in several branches of physics, as, for
example, in the description of spin chains, squeezed states, rotational
and vibrational nuclear and molecular spectra, and in conformal
field theories. By now several kinds of generalized deformed oscillators
$^{9-13}$
and generalized deformed su(2) algebras $^{14-20}$  have been introduced.

Here we shall confine ourselves to applications of quantum algebras in nuclear
structure physics. A brief description will be given of the su$_q$(2) rotator
model $^{21-26}$ and its extensions $^{14,27}$,  of the use of deformed
oscillators in the description of pairing correlations $^{28-30}$,
and of the formulation of deformed exactly soluble
models (Interacting Boson Model $^{31}$, Moszkowski model $^{32-34}$).
The purpose of this
short review is to provide the  reader with references for further reading.
Subsequently, the symmetries of the anisotropic quantum harmonic oscillator
with rational ratios of frequencies will be considered in more detail,
since they are of current interest $^{35,36}$ in connection with
superdeformed and hyperdeformed nuclei $^{37,38}$,
$\alpha$-cluster configurations in light nuclei $^{39-41}$,
and possibly with deformed atomic clusters $^{42,43}$.

\section{ The su$_q$(2) rotator model}

 The first application of quantum algebras in nuclear physics was the use
of the deformed algebra su$_q$(2) for the description of the rotational
spectra of deformed $^{21,22}$ and superdeformed $^{23}$ nuclei.
The same technique has been
used for the description of rotational spectra of diatomic molecules $^{24}$.
The Hamiltonian of the $q$-deformed rotator is proportional to the
second order Casimir operator of the su$_q$(2) algebra. Its Taylor expansion
contains powers of $J(J+1)$ (where $J$ is the angular momentum), being
similar to the expansion provided by the Variable Moment of Inertia (VMI)
model. Furthermore, the deformation
parameter $\tau$ (with $q=e^{i\tau}$) has been found to correspond to
the softness parameter of the VMI model $^{22}$.

B(E2) transition probabilities have also been described in this framework
$^{25}$.
In this case the $q$-deformed Clebsch--Gordan coefficients are used instead
of the normal ones. (It should be noticed that the $q$-deformed angular
momentum theory has already been much developed $^{25}$.) The model predicts
an increase of the B(E2) values with angular momentum, while the rigid
rotator model predicts saturation. Some experimental results supporting
this prediction already exist $^{25}$.

\section{ Extensions of the su$_q$(2) model}

The su$_q$(2) model has been successful in describing rotational nuclear
spectra. For the description of vibrational and transitional nuclear
spectra it has been found $^{27}$ that $J(J+1)$ has to be replaced by $J(J+c)$.
The additional parameter $c$ allows for the description of nuclear
anharmonicities in a way similar to that of the Interacting Boson Model
(IBM) $^{44,45}$ and the Generalized Variable Moment of Inertia (GVMI) model
$^{46}$.

Another generalization is based on the use of the deformed algebra
su$_{\Phi}$(2) $^{14}$, which is characterized by a structure function $\Phi$.
The usual su(2) and su$_q$(2) algebras are obtained for specific choices
of the structure function $\Phi$. The su$_{\Phi}$(2) algebra has been
constructed so that its representation theory resembles as much as possible
the representation theory of the usual su(2) algebra. Using this technique
one can construct, for example, a rotator having the same spectrum as the
one given by the Holmberg--Lipas formula $^{47}$. In addition to the
generalized
deformed su(2) algebra, generalized deformed oscillators $^{9-13}$
have also been
introduced and found useful in many physical applications.

\section{ Pairing correlations}

It has been found $^{28}$ that correlated fermion pairs coupled to zero
angular
momentum in a single-$j$ shell behave approximately as suitably defined
$q$-deformed bosons. After performing the same boson mapping to a simple
pairing Hamiltonian, one sees that the pairing energies are also correctly
reproduced up to the same order. The deformation parameter used ($\tau
=\ln q$) is found to be inversely proportional to the size of the shell,
thus serving as a small parameter.

The above mentioned system of correlated fermion pairs can be described
{\sl exactly} by suitably defined generalized deformed bosons $^{29}$. Then
both the commutation relations are satisfied exactly and the pairing energies
are reproduced exactly. The spectrum of the appropriate generalized
deformed oscillator corresponds, up to first order perturbation theory,
to a harmonic oscillator with an $x^4$ perturbation.

\section{ $q$-deformed versions of nuclear models}

A $q$-deformed version of a two dimensional toy Interacting Boson Model
(IBM) has been developed $^{31}$, mainly for testing the ways in which
spectra and transition probabilities are influenced by the $q$-deformation. A
$q$-deformed version of the full IBM is under development, while a
$q$-deformed version of the vibron model, which uses the IBM techniques
in the case of molecules, has already been developed $^{48}$.

Furthermore a $q$-deformed version of the Moszkowski model has been developed
$^{32}$ and RPA modes have been studied $^{33}$ in it. A $q$-deformed
Moszkowski model
with cranking has also been studied $^{34}$ in the mean-field approximation.
It has been seen that the residual interaction simulated by the
$q$-deformation is felt more strongly by states with large $J_z$. The
possibility of using $q$-deformation in assimilating temperature effects is
under discussion.

\section{ Anisotropic quantum harmonic oscillator with rational ratios of
frequencies}

The symmetries of the 3-dimensional anisotropic quantum harmonic oscillator
with rational ratios of frequencies (RHO) are of high current interest in
nuclear physics, since they are
the basic symmetries $^{35,36}$ underlying the structure of superdeformed and
hyperdeformed nuclei $^{37,38}$.
The 2-dimensional RHO is also of interest, in
connection with ``pancake'' nuclei, i.e. very oblate nuclei $^{36}$. Cluster
configurations in light nuclei can also be  described in terms of RHO
symmetries $^{39,40}$, which underlie the geometrical structure of the
Bloch--Brink $\alpha$-cluster model $^{41}$. The 3-dim RHO is also of
interest for the interpretation
of the observed shell structure in atomic clusters $^{42}$, especially
after the
realization $^{43}$ that large deformations can occur in such systems.

The two-dimensional $^{49-54}$ and
three-dimensional $^{55-61}$  anisotropic harmonic
oscillators have been the subject of several investigations, both at the
classical and the quantum mechanical level. These oscillators are examples
of superintegrable systems $^{62,63}$. The special cases with frequency
ratios 1:2 $^{64,65}$ and 1:3 $^{66}$ have also been considered. While
at the classical level it is clear that the su(N) or sp(2N,R) algebras can
be used for the description of the N-dimensional anisotropic oscillator, the
situation at the quantum level, even in the two-dimensional case, is not as
simple. The symmetry algebra of the two-dimensional anisotropic quantum
harmonic oscillator with rational ratio of frequencies has been identified as
a deformation of the u(2) algebra $^{67}$. The finite dimensional
representation modules of the algebra have been studied and the energy
eigenvalues have been determined using algebraic methods of general
applicability to quantum superintegrable systems $^{68}$. For labelling the
degenerate states an ``angular momentum'' operator has been introduced, the
eigenvalues of which are roots of appropriate generalized Hermite polynomials
$^{67,69,70}$. The cases with frequency ratios 1:$n$ have been found to
correspond to generalized parafermionic oscillators $^{71}$, while
in the special case with frequency ratio 2:1 the resulting algebra has been
found to correspond to the finite W algebra W$_3^{(2)}$ $^{72-75}$. For
more details the reader is referred to $^{67}$.

In this section  we are going to prove that a generalized deformed u(3)
algebra is the symmetry algebra of the three-dimensional anisotropic quantum
harmonic oscillator, which is the oscillator describing the single-particle
level spectrum of superdeformed and hyperdeformed nuclei.

\subsection{ The deformed u(N) algebra for the N-dimensional oscillator}

Let us consider the system described by the Hamiltonian:
\begin{equation}
H=\frac{1}{2}
\sum\limits_{k=1}^N \left(
{p_k}^2 + \frac{x_k^2}{m_k^2}
 \right),
\label{eq:Hamiltonian}
\end{equation}
where $m_i$ are natural numbers mutually prime ones, i.e.
their great common divisor is $\gcd (m_i,m_j)=1$ for $i\ne j$ and
$i,j=1,\ldots,N$.

We define the creation and annihilation operators $^{49}$:
\begin{equation}
\begin{array}{ll} a_k^\dagger=& \frac{1}{\sqrt{2}}
\left(\frac{x_k}{m_k} -i {p_k}  \right), \\
a_k=&\frac{1}{\sqrt{2}}
\left(\frac{x_k}{m_k} +i {p_k}  \right), \\
U_k=&
\frac{1}{2}
\left(
{p_k}^2 + \frac{x_k^2}{m_k^2}
\right)= \frac{1}{2} \left\{ a_k, a_k^\dagger \right\},
\quad H= \sum\limits_{k=1}^N U_k.
\end{array}
\label{eq:operators}
\end{equation}

These operators satisfy the relations (the indices $k$ have been omitted):
\begin{equation}
\begin{array}{ll}
a^\dagger\, U &= \left( U- \frac{1}{m} \right) a^\dagger \quad {\rm or} \quad
[U, \left(a^\dagger\right)^m ]= \left( a^\dagger \right)^m, \\
a\, U &= \left( U+ \frac{1}{m} \right)a \quad {\rm or}
 \quad [U, \left(a\right)^m ]= -\left(a\right)^m, \\
a^\dagger a &= U-\frac{1}{2m}, \qquad
a a^\dagger =U + \frac{1}{2m} , \\
\left[ a, a^{\dagger} \right] &= \frac{1}{m} , \\
\left(a^\dagger\right)^m \left(a\right)^m &=F(m,U), \\
\left(a\right)^m \left(a^\dagger\right)^m &=F(m,U+1), \\
\end{array}
\label{eq:basic}
\end{equation}
where the function $F(m,x)$ is defined by:
\begin{equation}
F(m,x)=\prod\limits_{p=1}^m{\left( x - \frac{2p-1}{2m}\right)\,}.\label{eq:F}
\end{equation}

Using the above relations we can define the enveloping
 algebra ${\cal C}$, defined by the polynomial combinations of the
generators $\Big\{ 1, H, {\cal A}_k, {\cal A}_k^\dagger, U_k
\Big\}$ and $k=1,\ldots, N-1$,  where:
\begin{equation}
{\cal A}_k^\dagger= \left( a_k^\dagger\right)^{m_k} \left( a_N \right)^{m_N},
\qquad
{\cal A}_k= \left( a_k\right)^{m_k} \left( a_N ^\dagger\right)^{m_N}.
\end{equation}
These operators correspond to a multidimensional generalization of eq.
(\ref{eq:operators}):
\begin{equation}
\big[ H, {\cal A}_k \big]=0,\quad
\big[ H, {\cal A}_k^\dagger \big]=0,\quad
\big[H, {U}_k  \big]=0,\quad
k=1,\ldots, N-1.
\label{eq:super}
\end{equation}
The following relations are satisfied for $k\ne \ell$
and $k,\ell =1,\ldots,N-1$:
\begin{equation}
\left[{U}_k, {\cal A}_\ell \right]=
\left[{U}_k, {\cal A}_\ell^\dagger \right]=
\left[{\cal A}_k, {\cal A}_\ell \right]=
\left[{\cal A}_k^\dagger, {\cal A}_\ell^\dagger \right]=0,
\label{eq:different}
\end{equation}
while
\begin{equation}
\begin{array}{l}
{U}_k{\cal A}_k^\dagger=  {\cal A}_k^\dagger \left( {U}_k +1 \right), \\
{U}_k {\cal A}_k=  {\cal A}_k\left( {U}_k -1 \right), \\
{\cal A}_k^\dagger{\cal A}_k=F\Big( m_k, {U}_k \Big) F\Big(m_N,  H
- \sum\limits_{\ell=1}^{N-1}{U}_\ell\, +1\Big),  \\
{\cal A}_k{\cal A}_k^\dagger=F\Big( m_k,   {U}_k +1 \Big) F\Big(m_N,  H
- \sum\limits_{\ell=1}^{N-1}{U}_\ell\Big). \\
\end{array}
\label{eq:kk}
\end{equation}
One  additional relation for $k\ne \ell$ can be derived:
\begin{equation}
F\Big(m_N,  H  - \sum\limits_{\ell=1}^{N-1}{U}_\ell\Big) {\cal A}_k^\dagger
{\cal A}_\ell =
F\Big(m_N,  H  - \sum\limits_{\ell=1}^{N-1}{U}_\ell\, +1\Big)
{\cal A}_\ell {\cal A}_k^\dagger.
\label{eq:kl}
\end{equation}

For illustrative purposes it is useful to work out the algebra in the case
of the 3-dim oscillator. Starting from the above definitions one ends up
with the commutation relations shown in Table 1, where
\begin{equation}
 {\cal A}_3^{\dagger} = (a_1^{\dagger})^{m_1} (a_2)^{m_2}, \qquad
   {\cal A}_3           = (a_1)^{m_1} (a_2^{\dagger})^{m_2},
\end{equation}
\begin{equation}
 F_i= F(m_i, U_i+1)-F(m_i, U_i),\qquad i=1,2,3,
\end{equation}
\begin{equation}
 F_{ij}= F(m_i,U_i+1) F(m_j, U_j) -F(m_i,U_i) F(m_j,U_j+1), \qquad
i,j=1,2,3.
\end{equation}
In the 1:1:1 case one has $F_i=1$, $i=1,2,3$ and $F_{ij}=U_j-U_i$,
$i,j=1,2,3$, so that Table 1 gives the usual results for the u(3) algebra.

In the 1:1:2 case the modified commutators read
$$ [{\cal A}_1, {\cal A}_1^{\dagger}]= -2 U_1 U_3 + U_3^2 +{3\over 16},\qquad
   [{\cal A}_1, {\cal A}_2^{\dagger}]= -2 U_3 {\cal A}_3, $$
$$ [{\cal A}_2, {\cal A}_2^{\dagger}]= -2 U_2 U_3 + U_3^2 +{3\over 16},\qquad
   [{\cal A}_2, {\cal A}_1^{\dagger}]= -2 U_3 {\cal A}_3^{\dagger},$$
Similarly in the 1:1:3 case one has
$$ [{\cal A}_1, {\cal A}_1^{\dagger}]= -3 U_1 U_3^2-{5\over 36} U_1 + U_3^3
+{23\over 36} U_3,\qquad
   [{\cal A}_1, {\cal A}_2^{\dagger}]= -\left(3 U_3^2+{5\over 36}\right)
{\cal A}_3, $$

\newpage

\centerline{\bf Table 1}

Commutation relations for the deformed u(3) algebra associated with the
$m_1:m_2:m_3$ case. The relevant operators are
defined in eqs. (2), (5), (10)--(12).

$$\vbox{\halign{\hfil #\hfil &&\quad \hfil #\hfil \cr
  & $U_1$ & $U_2$ & $U_3$ & ${\cal A}_3^{\dagger}$ & ${\cal A}_2^{\dagger}$ &
${\cal A}_1^{\dagger}$ & ${\cal A}_3$ & ${\cal A}_2$ & ${\cal A}_1$ \cr
$U_1$ & 0 & 0 & 0 & ${\cal A}_3^{\dagger}$ & 0 & ${\cal A}_1^{\dagger}$ &
$-{\cal A}_3$ & 0 & $-{\cal A}_1$ \cr
$U_2$ & 0 & 0 & 0 & $-{\cal A}_3^{\dagger}$ & ${\cal A}_2^{\dagger}$ & 0 &
${\cal A}_3$ & $-{\cal A}_2$ & 0 \cr
$U_3$ & 0 & 0 & 0 & 0 & $-{\cal A}_2^{\dagger}$ & $-{\cal A}_1^{\dagger}$ &
0 & ${\cal A}_2$ & ${\cal A}_1$ \cr
${\cal A}_3^{\dagger}$ & $-{\cal A}_3^{\dagger}$ & ${\cal A}_3^{\dagger}$ &
0 & 0 & ${\cal A}_1^{\dagger} F_2$ & 0 & $-F_{12}$ & 0 & $-{\cal A}_2 F_1$\cr
${\cal A}_2^{\dagger}$ & 0 & $-{\cal A}_2^{\dagger}$ & ${\cal A}_2^{\dagger}$
& $-{\cal A}_1^{\dagger}$ & 0 & 0 & 0 & $-F_{23}$ & ${\cal A}_3 F_3$ \cr
${\cal A}_1^{\dagger}$ & $-{\cal A}_1^{\dagger}$ & 0 & ${\cal A}_1^{\dagger}$
& 0 & 0 & 0 & $-{\cal A}_2^{\dagger} F_1$ & ${\cal A}_3^{\dagger} F_3$ &
$-F_{13}$ \cr
${\cal A}_3$ & ${\cal A}_3$ & $-{\cal A}_3$ & 0 & $F_{12}$ & 0 &
${\cal A}_2^{\dagger} F_1$ & 0 & $-{\cal A}_1 F_2 $ & 0 \cr
${\cal A}_2$ & 0 & ${\cal A}_2$ & $-{\cal A}_2$ & 0 & $F_{23}$ &
$-{\cal A}_3^{\dagger} F_3$ & ${\cal A}_1 F_2$ & 0 & 0\cr
${\cal A}_1$ & ${\cal A}_1$ & 0 & $-{\cal A}_1$ & ${\cal A}_2 F_1$ &
$-{\cal A}_3 F_3$ & $F_{13}$ & 0 & 0 & 0 \cr}}$$

\hrule

$$ [{\cal A}_2, {\cal A}_2^{\dagger}]= -3 U_2 U_3^2-{5\over 36} U_2 +U_3^3
+{23\over 36} U_3, \qquad
   [{\cal A}_2, {\cal A}_1^{\dagger}]= -\left(3 U_3^2+{5\over 36}\right)
{\cal A}_3^{\dagger}.$$
In the same way in the 1:1:4 case one has
$$ [{\cal A}_1, {\cal A}_1^{\dagger}]= -4 U_1 U_3^3- {11\over 16} U_1 U_3+
U_3^4 +{43\over 32} U_3^2+{105\over 4096}, \qquad
   [{\cal A}_1, {\cal A}_2^{\dagger}]= -\left( 4 U_3^3 +{11\over 16} U_3\right)
{\cal A}_3,$$
$$ [{\cal A}_2, {\cal A}_2^{\dagger}]= -4 U_2 U_3^3 -{11\over 16} U_2 U_3+
U_3^4 +{43\over 32} U_3^2 +{105\over 4096}, \qquad
   [{\cal A}_2, {\cal A}_1^{\dagger}]= -\left( 4 U_3^3+{11\over 16} U_3\right)
{\cal A}_3^{\dagger}.$$
We remark that for the 1:1:m oscillators only four commutators get modified
in each case.

In the 2:2:1 case the modified commutators read
$$ [{\cal A}_1, {\cal A}_1^{\dagger}]= 2 U_1 U_3 -U_1^2-{3\over 16}, \qquad
   [{\cal A}_1, {\cal A}_3^{\dagger}]= 2 U_1 {\cal A}_2, $$
$$ [{\cal A}_2, {\cal A}_2^{\dagger}]= 2 U_2 U_3 -U_2^2-{3\over 16},\qquad
   [{\cal A}_2, {\cal A}_3^{\dagger}]= 2 U_2 {\cal A}_1, $$
$$ [{\cal A}_3, {\cal A}_3^{\dagger}]= -(U_1-U_2)\left( 2 U_1 U_2 -{3\over 8}
\right) ,\qquad
   [{\cal A}_3, {\cal A}_1^{\dagger}]= 2 U_1 {\cal A}_2^{\dagger},$$
$$ [{\cal A}_3, {\cal A}_2^{\dagger}]= 2 U_2
{\cal A}_1^{\dagger},$$
while the commutators
$$ [{\cal A}_1, {\cal A}_2^{\dagger}]= -{\cal A}_3, \qquad
   [{\cal A}_2, {\cal A}_1^{\dagger}]= -{\cal A}_3^{\dagger}, $$
are  not modified in this case. One can see that for the m:m:1 oscillators
seven commutators get modified in each case.

\subsection{ The representations of the deformed u(N)}

The algebra defined in the previous subsection accepts a Fock space
 representation. The elements of the basis
$\left\vert E, p_1,\ldots,p_{N-1} \right>$
are characterized by the
eigenvalues of the $N$
commuting elements  of the algebra $H$ and ${U}_k$ with
$k=1,\ldots,N-1$. The elements ${\cal A}_k$ and ${\cal A}_k^\dagger$
are the corresponding ladder operators of the algebra. The following
relations hold:

\begin{equation}
\begin{array}{l}
H \left\vert E, p_1,\ldots,p_{N-1} \right>=
E\left\vert E, p_1,\ldots,p_{N-1} \right>, \\
{U}_k \left\vert E, p_1,\ldots,p_{N-1} \right>=
p_k\left\vert E, p_1,\ldots,p_{N-1} \right>, \\
{\cal A}_k \left\vert E, p_1,\ldots,p_k,\ldots,p_{N-1} \right>=\\
\sqrt{F\Big( m_k,   { p}_k\Big) F\Big(m_N,  E
- \sum\limits_{\ell=1}^{N-1}{p}_\ell+1\Big) }
\left\vert E, p_1,\ldots,p_k-1,\ldots,p_{N-1} \right>, \\
{\cal A}_k^\dagger \left\vert E, p_1,\ldots,p_k,\ldots,p_{N-1} \right>=\\
\sqrt{F\Big( m_k,   { p}_k+1 \Big) F\Big(m_N,  E
- \sum\limits_{\ell=1}^{N-1}{p}_\ell\Big)}
\left\vert E, p_1,\ldots,p_k+1,\ldots,p_{N-1} \right>. \\
\end{array}
\label{eq:repr}
\end{equation}

Let $p_k^{\mbox{min}}$ be the minimum value of $p_k$ such that
\begin{equation}
{\cal A}_k \left\vert E, p_1,\ldots,p_k^{\mbox{min}},\ldots,p_{N-1}
 \right>=0.
\label{eq:maxweight}
\end{equation}
{}From eq. (\ref{eq:repr}) we find that we must have:
\begin{equation}
F\Big( m_k,{ p}_k^{\mbox{min}} \Big)=0. \label{eq:1}
\end{equation}
Then
 $p_k$ is one of the roots of the function $F$ defined by
eq. (\ref{eq:F}).
The general form of the roots is:
\begin{equation}
p_k^{\mbox{min}}=\frac{2q_k -1 }{2 m_k }, \qquad q_k= 1,\ldots,m_k.
\label{eq:roots}
\end{equation}
Each root is characterized by a number $q_k$.  The numbers $q_k$
 also characterize the representations of the algebra, as we shall
see.

The elements of the Fock space can be generated by successive
applications of the ladder operators ${\cal A}_k^\dagger$ on the
minimum weight element
\begin{equation}
\left\vert E, p_1^{\mbox{min}},\ldots, p_{N-1}^{\mbox{min}}\right>=\left\vert
\begin{array}{c} E, \left[0\right]\\ \left[ q \right] \end{array}
\right>=\left\vert \begin{array}{c} E, 0, \ldots , 0 \\ q_1,
\ldots,q_k,\ldots,q_{N} \end{array} \right>.
\label{eq:min}
\end{equation}
The elements of the basis of the Fock space are given by:
\begin{equation}
\left\vert E, [p_k^{\mbox{min}}+n_k] \right>=
\left\vert \begin{array}{c} E, \left[n\right]\\ \left[ q \right] \end{array}
\right>
= \frac{1}{\sqrt{C^{[n]}_{[q]}}} \left(
\prod\limits_{k=1}^{N-1}\left({\cal A}_k^\dagger\right)^{n_k} \right)
\left\vert\begin{array}{c} E, \left[0\right]\\  \left[ q \right]  \end{array}
\right>,
\label{eq:base}
\end{equation}
where $[n]= (n_1, n_2, \cdots, n_{N-1})$ and $[q]= (q_1,q_2, \ldots, q_{N})$,
while $C^{[n]}_{[q]}$ are normalization coefficients.

The generators of the algebra acting on the base  of the Fock space give:
\begin{equation}
\begin{array}{l}
H \left\vert\begin{array}{c}E, \left[n\right]\\ \left[ q \right] \end{array}
\right>=E \left\vert \begin{array}{c} E, \left[n\right]\\ \left[ q \right]
\end{array} \right>, \\
{U}_k \left\vert \begin{array}{c} E, \left[n\right]\\ \left[ q \right]
\end{array} \right> = \left(n_k + p_k^{\mbox{min}}\right) \left\vert
\begin{array}{c} E, \left[n\right]\\ \left[ q \right] \end{array} \right>,
\quad k=1,\ldots,N-1, \\ \end{array}
\label{eq:diag}
\end{equation}
\begin{equation}
\begin{array}{l}
{\cal A}^\dagger_k
\left\vert
\begin{array}{c}
E, \left[n\right]\\
\left[ q \right]
\end{array}
\right>
=\\
=
\sqrt{
F\Big( m_k,  n_k+p_k^{\mbox{min}} +1 \Big)
F\Big(m_N,  E
- \sum\limits_{\ell=1}^{N-1}\left(n_\ell+p_\ell^{\mbox{min}}\right)\Big)
}
\cdot\\
\quad \cdot
\left\vert
\begin{array}{c}
E, n_1,\ldots, n_k+1,\ldots,n_{N-1}\\
 q_1,\ldots,q_k,\ldots,
q_N
\end{array}
\right>,
\end{array}
\label{eq:Ak+}
\end{equation}
\begin{equation}
\begin{array}{l}
{\cal A}_k
\left\vert
\begin{array}{c}
E, \left[n\right]\\
\left[ q \right]
\end{array}
\right>
=\\
=
\sqrt{
F\Big( m_k,   n_k+p_k^{\mbox{min}} \Big)
F\Big(m_N,  E
- \sum\limits_{\ell=1}^{N-1}\left(n_\ell+p_\ell^{\mbox{min}}\right)+1\Big)
}
\cdot\\
\quad \cdot
\left\vert
\begin{array}{c}
E, n_1,\ldots, n_k-1,\ldots,n_{N-1}\\
 q_1,\ldots,q_k,\ldots,
q_N
\end{array}
\right>.
\end{array}
\label{eq:Ak}
\end{equation}
The existence of a finite dimensional representation implies
that after $\Sigma$ successive applications of the ladder
operators ${\cal A}^\dagger$ on the minimum weight element one gets zero,
so that the following condition is satisfied:
$$ F(m_N,E-\Sigma - \sum\limits_{\ell=1}^{N-1}p_\ell^{\mbox{min}})=0. $$
Therefore:
$$E -\Sigma - \sum\limits_{\ell=1}^{N-1}p_\ell^{\mbox{min}}=p_N^{\mbox{min}},$$
where $p_N^{\mbox{min}}$ is the root of equation
$F(m_N,p_N^{\mbox{min}})=0$. Then
\begin{equation}
E= \Sigma + \sum\limits_{k=1}^{N} \frac{2q_k-1}{2 m_k}\label{eq:energy}.
\end{equation}
In the case of finite dimensional representations  only the energies given
by eq.  (\ref{eq:energy}) are permitted and the elements of the
Fock space can be described by using $\Sigma$ instead of  $E$.
The action of the generators on the Fock space is described by
the following relations:
\begin{equation}
\begin{array}{l}
H\left\vert \begin{array}{c} \Sigma, \left[n\right]\\ \left[ q \right]
\end{array} \right>=  \left( \Sigma + \sum\limits_{k=1}^{N}
\frac{2q_k-1}{2 m_k} \right) \left\vert \begin{array}{c} \Sigma,
\left[n\right]\\ \left[ q \right] \end{array} \right>, \\
{U}_k \left\vert \begin{array}{c} \Sigma, \left[n\right]\\
\left[ q \right]
\end{array}
\right>
=
\left(n_k + \frac{2q_k-1}{2 m_k}\right)
\left\vert
\begin{array}{c}
\Sigma, \left[n\right]\\
\left[ q \right]
\end{array}
\right>, \quad k=1,\ldots,N-1, \\
\end{array}
\label{eq:diag1}
\end{equation}
\begin{equation}
\begin{array}{l}
{\cal A}^\dagger_k
\left\vert
\begin{array}{c}
\Sigma, \left[n\right]\\
\left[ q \right]
\end{array}
\right>
=\\
=
\sqrt{
F\Big( m_k,   n_k
+\frac{2q_k-1}{2 m_k} +1 \Big)
F\Big(m_N,  \Sigma
- \sum\limits_{\ell=1}^{N-1}n_\ell\,
+\frac{2q_N-1}{2 m_N}\Big)
}
\cdot\\
\quad \cdot
\left\vert
\begin{array}{c}
\Sigma, n_1,\ldots, n_k+1,\ldots,n_{N-1}\\
 q_1,\ldots,q_k,\ldots,
q_N
\end{array}
\right>,
\end{array}
\label{eq:Ak1+}
\end{equation}
\begin{equation}
\begin{array}{l}
{\cal A}_k
\left\vert
\begin{array}{c}
\Sigma, \left[n\right]\\
\left[ q \right]
\end{array}
\right>
=\\
=
\sqrt{
F\Big( m_k,   n_k+\frac{2q_k-1}{2 m_k} \Big)
F\Big(m_N,  \Sigma
- \sum\limits_{\ell=1}^{N-1}n_\ell\,+
\frac{2q_N-1}{2 m_N}+1\Big)
}
\cdot\\
\quad \cdot
\left\vert
\begin{array}{c}
\Sigma, n_1,\ldots, n_k-1,\ldots,n_{N-1}\\
 q_1,\ldots,q_k,\ldots,
q_N
\end{array}
\right>.
\end{array}
\label{eq:Ak1}
\end{equation}
The dimension of the representation is given by
$$ d= {\Sigma+N-1 \choose \Sigma }= {(\Sigma+1) (\Sigma+2)\cdots (\Sigma+N-1)
\over (N-1)!}.$$
It is clear that to each value of $\Sigma$ correspond $m_1 m_2 \dots m_N$
energy eigenvalues, each eigenvalue having degeneracy $d$.

\subsection{ Connection to the Cartesian basis}

Using eqs (\ref{eq:basic}) we can prove that
the algebra  generated by the generators
$a_\ell^\dagger,a_\ell, N_\ell=m_\ell U_\ell-1/2$ is an oscillator algebra
with structure function $^{9,12}$
 $$\Phi_\ell(x)= x/m_\ell, $$
i.e. with
$$ \Phi_l (N_l)= U_l-{1\over 2 m_l}.$$
This oscillator algebra is characterized by the commutation relations:
\begin{equation}
\left[ N_\ell, a_\ell^\dagger\right]=a^\dagger_\ell,
\quad
\left[ N_\ell, a_\ell\right]=- a_\ell,
\quad
a^\dagger_\ell a_\ell = \Phi_\ell\left( N_\ell \right),
\quad
a_\ell a_\ell^\dagger = \Phi_\ell\left( N_\ell +1 \right).
\label{eq:oscalgebra}
\end{equation}
There are in total $N$ different oscillators of this type, uncoupled
to each other. The  Fock space corresponding
 to these oscillators defines an infinite dimensional representation of
the algebra defined eqs (\ref{eq:super}-\ref{eq:kl}).
In order to see the connection of the present basis to the usual Cartesian
basis, one can use for the latter the symbol $[r]=(r_1,r_2,\ldots,r_N)$.
One then has
$$
\begin{array}{l}
a^\dagger_\ell
\left\vert [r] \right>=
\sqrt{\Phi(r_\ell+1)} \left\vert r_1,\ldots,r_\ell+1,\ldots,r_N \right>,
\\
a_\ell
\left\vert [r] \right>=
\sqrt{\Phi(r_\ell)} \left\vert r_1,\ldots,r_\ell-1,\ldots,r_N \right>,
\\
N_\ell
\left\vert [r] \right>=
r_\ell\left\vert [r] \right>.
\end{array}
$$
The connection between the above basis and the basis defined by eqs
(\ref{eq:diag1}-\ref{eq:Ak1}) is given by:
\begin{equation}
\begin{array}{c}
\left\vert [r] \right>=
\left\vert
\begin{array}{c}
\Sigma, \left[n\right]\\
\left[ q \right]
\end{array}
\right>, \\
\begin{array}{c}
r_\ell=n_\ell m_\ell + \mbox{mod}\left(r_\ell,m_\ell\right)\\
\ell = 1,\ldots, N
\end{array}
\longleftrightarrow
\begin{array}{l}
n_k=\left[ r_k/m_k \right]\\
k=1,\ldots,N-1\\
q_\ell=\mbox{mod }\left(r_\ell,m_\ell\right)+1\\
\Sigma = \sum\limits_{\ell=1}^N \left[ r_\ell/m_\ell\right]
\end{array}
\end{array}
\label{eq:corresp}
\end{equation}
where $[x]$ means the integer part of the number $x$.

Using the correspondence between the present basis and the usual Cartesian
basis, given in eq. (\ref{eq:corresp}), the action  of the operators
$a_k^\dagger$ on the present basis can be
calculated for $k=1,\ldots,N-1$:
$$
a^\dagger_k
\left\vert
\begin{array}{c}
\Sigma, \left[n\right]\\
\left[ q \right]
\end{array}
\right>
=
\sqrt{ n_k + q_k/m_k }
\left\vert
\begin{array}{c}
\Sigma', \left[n'\right]\\
\left[ q' \right]
\end{array}
\right>,
$$
where
$$
\begin{array}{c}
n'_\ell=n_\ell \quad q'_\ell = q_\ell \quad  \mbox{for } \ell \ne k, \\
n'_k = n_k + \left[ q_k/m_k\right], \\
\Sigma' = \Sigma + \left[ q_k/m_k\right], \\
q'_k= \mbox{mod }\left(q_k, m_k \right) +1,
\end{array}
$$
while for the operator $a^\dagger_N$ one has
$$
a^\dagger_N
\left\vert
\begin{array}{c}
\Sigma, \left[n\right]\\
\left[ q \right]
\end{array}
\right>
=
\sqrt{ \Sigma - \sum\limits_{k=1}^{N-1}n_k\, + q_N/m_N }
\left\vert
\begin{array}{c}
\Sigma', \left[n'\right]\\
\left[ q' \right]
\end{array}
\right>,
$$
where
$$
\begin{array}{c}
n'_k=n_k,  \qquad q'_k = q_k,  \qquad  \mbox{for }  k=1,\ldots,N-1, \\
\Sigma' = \Sigma + \left[ q_N/m_N\right], \\
q'_N= \mbox{mod }\left(q_N, m_N \right) +1.
\end{array}
$$
Similarly for the operators $a_k$ one can find for $k=1,$ \dots, $N-1$:
$$
a_k
\left\vert
\begin{array}{c}
\Sigma, \left[n\right]\\
\left[ q \right]
\end{array}
\right>
=
\sqrt{ n_k + (q_k-1)/m_k }
\left\vert
\begin{array}{c}
\Sigma', \left[n'\right]\\
\left[ q' \right]
\end{array}
\right>,
$$
where
$$
\begin{array}{c}
n'_\ell=n_\ell,  \qquad q'_\ell = q_\ell,  \qquad  \mbox{for } \ell \ne k, \\
n'_k = n_k + \left[ (q_k-2)/m_k\right], \\
\Sigma' = \Sigma + \left[ (q_k-2)/m_k\right], \\
q'_k= \mbox{mod }\left(q_k-2, m_k \right) +1,
\end{array}
$$
while for the operator $a_N$ one has
$$
a_N
\left\vert
\begin{array}{c}
\Sigma, \left[n\right]\\
\left[ q \right]
\end{array}
\right>
=
\sqrt{ \Sigma - \sum\limits_{k=1}^{N-1}n_k\, + (q_N-1)/m_N }
\left\vert
\begin{array}{c} \Sigma', \left[n'\right]\\ \left[ q' \right] \end{array}
\right>,
$$
where
$$
\begin{array}{c}
n'_k=n_k,  \qquad q'_k = q_k,  \qquad  \mbox{for }  k=1,\ldots,N-1, \\
\Sigma' = \Sigma + \left[ (q_N-2)/m_N\right], \\
q'_N= \mbox{mod }\left(q_N-2, m_N \right) +1.
\end{array}
$$

For illustrative purposes, we shall discuss a few cases in some detail.

i) In the 1:1:1 case  (isotropic 3-dim oscillator) the only allowed
$[ q_1 q_2 q_3]$ set is $[1 1 1]$. In the Cartesian notation $| r_1 r_2 r_3>$
the lowest energy corresponds to the state $|000>$, the next energy level
corresponds to the three states $|100>$, $|010>$ and $|001>$, while the
next energy level corresponds to the six states $|200>$, $|020>$, $|002>$,
$|110>$, $|101>$, $|011>$. In the basis of eqs. (23)-(25) these states
are written as
$$ |000>= \left\vert\begin{array}{c} 0, 0 0 \\111\end{array}\right>, \qquad$$
$$   |100>= \left\vert\begin{array}{c} 1, 1 0 \\111\end{array}\right>,\qquad
   |010>= \left\vert\begin{array}{c} 1, 01 \\111 \end{array}\right>,\qquad
   |001>= \left\vert\begin{array}{c} 1, 00 \\111 \end{array}\right>,$$
$$ |200>=\left\vert\begin{array}{c} 2, 20 \\111\end{array}\right>,\qquad
   |020>=\left\vert\begin{array}{c} 2,02\\111\end{array}\right>,\qquad
   |002>=\left\vert\begin{array}{c} 2,00\\111\end{array}\right>,$$
$$ |110>=\left\vert\begin{array}{c} 2,11\\111\end{array}\right>,\qquad
   |101>=\left\vert\begin{array}{c} 2,10\\111\end{array}\right>,\qquad
   |011>=\left\vert\begin{array}{c} 2,01\\111\end{array}\right>.$$
It is clear that the irreps are characterized by the quantum numbers
$\Sigma$ and $[q_1 q_2 q_3]$, while $[n_1 n_2]$ enumerate the degenerate
states within each irrep. The lowest irrep, characerized by $\Sigma=0$ and
$[q_1 q_2 q_3]=[111]$, has dimension $d=1$, the next irrep is characterized
by $\Sigma=1$ and $[q_1 q_2 q_3]=[111]$ and has dimension $d=3$, while
the next irrep has $\Sigma=2$, $[q_1q_2q_3]=[111]$ and dimension $d=6$.
According to eq. (26) the dimensions of the first few u(3)  irreps are
1, 3, 6, 10, 15, 21, \dots.

ii) In the 1:1:2 case the allowed $[q_1 q_2 q_3]$ sets are $[111]$ and
$[112]$. The lowest irrep is characterized by $\Sigma=0$, $[q_1q_2q_3]
=[111]$, has dimension 1 and contains the state $|000>$. The second
irrep has $\Sigma=0$, $[q_1q_2q_3]=[112]$, $d=1$ and contains the state
$|001>$. The third irrep has $\Sigma=1$, $[q_1q_2q_3]=[111]$, $d=3$, and
contains the states $|100>$, $|010>$, $|002>$. The fourth irrep is
characterized by $\Sigma=1$, $[q_1q_2q_3]=[112]$, has dimension $d=3$ and
contains the states $|101>$, $|011>$, $|003>$. The states are listed
in both notations below:
$$ |000>=\left\vert\begin{array}{c} 0,00\\111\end{array}\right>,\qquad
   |001>=\left\vert\begin{array}{c} 0,00\\112\end{array}\right>,$$
$$ |100>=\left\vert\begin{array}{c} 1,10\\111\end{array}\right>, \qquad
   |010>=\left\vert\begin{array}{c} 1,01\\111\end{array}\right>, \qquad
   |002>=\left\vert\begin{array}{c} 1,00\\111\end{array}\right>,$$
$$ |101>=\left\vert\begin{array}{c} 1,10\\112\end{array}\right>,\qquad
   |011>=\left\vert\begin{array}{c} 1,01\\112\end{array}\right>,\qquad
   |003>=\left\vert\begin{array}{c} 1,00\\112\end{array}\right>.$$

iii) In the 2:2:1 case the allowed $[q_1q_2q_3]$ sets are $[111]$, $[211]$,
$[121]$, $[221]$. The lowest energy level is characterized by $\Sigma=0$,
$[q_1q_2q_3]=[111]$, has dimension $d=1$ and contains the state $|000>$.
The next energy level has $d=2$ and is containing the 1-dim irrep with
$\Sigma=0$, $[q_1q_2q_3]=[211]$, i.e. the state $|100>$, and the
1-dim irrep with $\Sigma=0$, $[q_1q_2q_3]=[121]$, i.e. the state $|010>$.
The next energy level has $d=4$ and contains two irreps: the 1-dim irrep
with $\Sigma=0$, $[q_1q_2q_3]=[221]$ (state $|110>$), and the 3-dim irrep
with $\Sigma=1$, $[q_1q_2q_3]=[111]$ (states $|001>$, $|200>$, $|020>$).
The next energy level has $d=6$ and contains two irreps: the 3-dim irrep
with $\Sigma=1$, $[q_1q_2q_3]=[211]$ (states $|101>$, $|300>$, $|120>$),
and the 3-dim irrep with $\Sigma=1$, $[q_1q_2q_3]=[121]$ (states $|011>$,
$|030>$, $|210>$). These states are listed in both notations below:
$$ |000>=\left\vert\begin{array}{c} 0,00\\111\end{array}\right>,\qquad
   |100>=\left\vert\begin{array}{c} 0,10\\211\end{array}\right>,\qquad
   |010>=\left\vert\begin{array}{c} 0,01\\121\end{array}\right>,  $$
$$ |110>=\left\vert\begin{array}{c} 0,11\\221\end{array}\right>,\qquad
   |001>=\left\vert\begin{array}{c} 1,00\\111\end{array}\right>,\qquad $$
$$   |200>=\left\vert\begin{array}{c} 1,20\\111\end{array}\right>,\qquad
   |020>=\left\vert\begin{array}{c} 1,02\\111\end{array}\right>,$$
$$ |101>=\left\vert\begin{array}{c} 1,10\\211\end{array}\right>,\qquad
   |300>=\left\vert\begin{array}{c} 1,30\\211\end{array}\right>,\qquad
   |120>=\left\vert\begin{array}{c} 1,12\\211\end{array}\right>,$$
$$ |011>=\left\vert\begin{array}{c} 1,01\\121\end{array}\right>,\qquad
   |030>=\left\vert\begin{array}{c} 1,03\\121\end{array}\right>,\qquad
   |210>=\left\vert\begin{array}{c} 1,21\\121\end{array}\right>.$$

The following comments can now be made:

i) In the basis described by eqs (23)-(25) it is a trivial matter to
distinguish the states belonging to the same irrep for any $m_1:m_2:m_3$
ratios, while in the Cartesian basis this is true only in the 1:1:1 case.

ii) In the 1:1:2 case we see that the irreps have degeneracies 1, 1, 3, 3,
6, 6, 10, 10, \dots, i.e. ``two copies'' of the u(3) degeneracies 1, 3,
6, 10, \dots are obtained.

iii) It can be easily seen that the 1:1:n case corresponds to ``n copies''
of the u(3) degeneracies. In the case 1:1:3,  for example, the degeneracies
are 1, 1, 1, 3, 3, 3, 6, 6, 6, \dots.

iv) In the 2:2:1 case the energy levels do not correspond to a single
irrep each, but some of them correspond to sums of irreps, i.e. to
reducible representations (rreps) of the deformed u(3) algebra.
The same is true for the m:m:1 case. In the 2:2:1 case, in particular,
the degeneracies are 1, 2, 4, 6, 9, 12, 16, \dots, which correspond
to the dimensions of the irreps of O(4) $^{76}$.

v) It can be easily seen that the condition for each energy eigenvalue
to correspond to one irrep of the deformed algebra is that $m_1$, $m_2$,
$m_3$ are mutually prime numbers. If two of them possess a common divisor
other than 1, then some energy eigenvalues will correspond to sums of
irreps, i.e. to rreps.

vi) Cases where rreps appear, can be approximated by cases where only
irreps appear. For example, 2:2:1 can be approximated by 21:19:10 or
201:199:100. (See $^{52}$ for more details.)

vii) The difference between the formalism used here and the one used in
$^{55,57,58,61}$ is that in the latter case for given $m_1$, $m_2$,
$m_3$, appropriate operators have to be introduced separately for each set
of $[q_1 q_2 q_3]$ values, while in the present case only one set of
operators is introduced.  It should be noticed that
$(s)$ of refs $^{55,57,58}$ ($\{\lambda\}$ of ref. $^{61}$) is analogous
to the $[q]$ used in the present work.

\subsection{ The 3-dimensional oscillator and relation to the  Nilsson
model}

In this subsection the 3-dim case will be studied in more detail, because of
its
relevance for the description of superdeformed nuclei and of nuclear and
atomic clusters. The 3-dim anisotropic oscillator is the basic ingredient of
the Nilsson model $^{77}$, which in addition contains a term proportional
to the square of the angular momentum operator, as well as a spin-orbit
coupling term, the relevant Hamiltonian being
$$ H_{Nilsson}= H_{osc} - 2 k \vec{L} \cdot \vec{S} - k \nu \vec{L}^2,$$
where $k$, $\nu$ are constants.
 The spin-orbit term is not needed in the case of atomic
clusters, while in the case of nuclei it can be effectively removed through
a unitary transformation, both in the case of the spherical Nilsson model
$^{78-80}$ and of the axially symmetric one $^{81,82}$.
An alternative way to effectively remove the spin-orbit term in the
spherical Nilsson model is the q-deformation of the relevant algebra
$^{83}$. It should also be noticed that the spherical Nilsson
Hamiltonian is known to possess an osp(1$\vert$2) supersymmetry $^{84}$.

In the case of the 3-dim oscillator the relevant operators of eq. (2)  form a
nonlinear generalization of the algebra u(3), the q-deformed version of
which can be found in $^{31,85,86}$.

As we have already seen, to each $\Sigma$ value correspond $m_1 m_2 m_3$
energy eigenvalues, each eigenvalue having degeneracy $(\Sigma+1)(\Sigma+2)/2$.
In order to distinguish the degenerate eigenvalues, we are going to introduce
some generalized angular momentum operators, $L_i$ ($i=1,2,3$),  defined by:
\begin{equation}
L_k =
i \epsilon_{ijk} \left(
\left( a_i \right)^{m_i} \left( a_j^\dagger \right)^{m_j}-
\left( a_i^\dagger \right)^{m_i} \left( a_j \right)^{m_j}
\right).
\label{eq:defang}
\end{equation}
One  can prove that
$$
L_1=i \left( {\cal A}_2 - {\cal A}_2^\dagger \right),
\quad
L_2=i \left( {\cal A}^\dagger_1 - {\cal A}_1 \right).
$$
The following commutation relations can be verified:
\begin{equation}
\left[ L_i, L_j \right] = i \epsilon_{ijk}
\left( F( m_k, U_k+1) - F( m_k, U_k ) \right)L_k.
\label{eq:angmom}
\end{equation}
It is worth noticing that in the case of $m_1=m_2=m_3=1$ one has
$F(1,x)=x-1/2$, so that the above equation gives the usual angular
momentum  commutation relations.

The operators defined in  eq. (\ref{eq:defang}) commute with the
oscillator  hamiltonian
$H$ and therefore conserve the number $\Sigma$ which characterizes the
dimension of the representation. Also these operators do not change the
numbers $q_1$, $q_2$, $q_3$ as we can see from eqs
(\ref{eq:Ak1+}-\ref{eq:Ak1}).
The eigenvalues of these operators can be calculated using Hermite
function techniques.

Let us consider in particular the generalized angular momentum projection:
\begin{equation}
L_3= i
\left( \left(a_1\right)^{m_1} \left(a_2^\dagger\right)^{m_2}-
\left(a_1^\dagger\right)^{m_1} \left(a_2\right)^{m_2} \right).
 \label{eq:l3}
\end{equation}
This acts on the basis as follows
\begin{equation}
\begin{array}{l}
L_3
\left\vert
\begin{array}{c}
\Sigma, \left[n\right]\\
\left[ q \right]
\end{array}
\right>
=\\
=i
\Big(
\sqrt{ F\left(m_1, n_1 + \frac{2 q_{1} -1}{2m_1} \right)
F\left(m_2, n_2 + \frac{2 q_{2} -1}{2m_2}+1 \right)}
\left\vert
\begin{array}{c}
\Sigma, n_{1}-1, n_{2}+1\\
\left[ q \right]
\end{array}
\right>
-\\
-
\sqrt{ F\left(m_1, n_1 + \frac{2 q_{1} -1}{2m_1} +1 \right)
F\left(m_2, n_2 + \frac{2 q_{2} -1}{2m_2} \right)}
\left\vert
\begin{array}{c}
\Sigma, n_{1}+1, n_{2}-1\\
\left[ q \right]
\end{array}
\right>\Big).
\end{array}
\label{eq:L3a}
\end{equation}
This operator conserves the quantum number
$$
j= \frac{n_1+n_2}{2}, \quad j=0,{1\over 2},1,{3\over 2},\ldots
$$
In addition one can  introduce the quantum number
$$
m ={n_1-n_2\over 2}.
$$
One can then replace the quantum numbers $n_1$, $n_2$ by the quantum
numbers $j$, $\mu$, where $\mu$ is the eigenvalue
of the $L_3$ operator. The new representation basis one can label as
\begin{equation}
L_3
\left\vert
\begin{array}{c}
\Sigma,\\
 j, \mu \\
\left[ q \right]
\end{array}
\right>
=
\mu
\left\vert
\begin{array}{c}
\Sigma,\\
 j, \mu \\
\left[ q \right]
\end{array}
\right>.
\label{eq:eigenL3}
\end{equation}
This basis is connected to the basis of the previous section as follows
\begin{equation}
\left\vert
\begin{array}{c}
\Sigma,\\
 j, \mu \\
\left[ q \right]
\end{array}
\right>
=
\sum\limits_{m=-j}^{j}
\frac{ c[j,m,\mu]}
{\sqrt{ [j+m]_1! [j-m]_2! }}
\left\vert
\begin{array}{c}
\Sigma, j+m, j-m\\
\left[ q \right]
\end{array}
\right>,
\label{eq:L3basis}
\end{equation}
where
$$
[0]_k!=1,\quad
[n]_k! = [n]_k [n-1]_k!,
\quad
[n]_k=F\left( m_k, n + \frac{2q_k-1}{2m_k} \right),
$$
and the coefficients  $c[j,m,\mu]$  in eq.
(\ref{eq:L3basis}) satisfy the recurrence relation:
\begin{equation}
\mu c[j,m,\mu] =i
\left(
 [j-m]_2 c[j,m+1,\mu] -[j+m]_1c[j,m-1,\mu]
\right).
\label{eq:L3rec}
\end{equation}
These relations can be satisfied only for special values of
the parameter $\mu$, corresponding to the eigenvalues of $L_3$.
It is worth noticing that in the case
 of $m_1=m_2$, which corresponds to axially symmetric oscillators,
the possible values turn out to be $\mu=-2j, -2(j-1), \ldots, 2(j-1),2j$.
In nuclear physics the quantum numbers $n_{\bot}=n_1+n_2$ and
$\Lambda = \pm n_{\bot}$, $\pm(n_{\bot}-2)$, \dots, $\pm 1$ or 0 are used
$^{87}$. From the above definitions it is clear that $j= n_{\bot}/2$ and
$\mu = \Lambda$. Therefore in the case of $m_1=m_2$, which includes
axially symmetric prolate nuclei with $m_1:m_2:m_3= 1:1:m$, as well as
axially symmetric oblate nuclei with $m_1:m_2:m_3= m:m:1$, the
correspondence between the present scheme and the Nilsson model is clear.


\subsection{  Summary}

The symmetry algebra of the N-dim anisotropic quantum harmonic oscillator
with rational ratios of frequencies has been constructed by a method
$^{68}$ of general applicability in constructing finite-dimensional
representations of quantum superinegrable systems.  The case of the
3-dim oscillator has been considered in detail, because of its relevance
to the single particle level spectrum of superdeformed and
hyperdeformed nuclei $^{35,36}$, to the underlying geometrical
structure of the Bloch-Brink $\alpha$-cluster model $^{39-41}$, and possibly
to the shell structure of atomic clusters at large deformations
$^{42,43}$.  The symmetry algebra in this case is a nonlinear
generalization of the u(3) algebra. For labeling the degenerate states,
generalized angular momentum operators are introduced, clarifying the
connection of the present approach to the Nilsson model.

In the case of the 2-dim oscillator with ratio of frequencies $2:1$
($m_1=1$, $m_2 =2$) it has been shown $^{67}$ that the relevant nonlinear
generalized u(2) algebra can be identified as the finite W algebra
W$_3^{(2)}$ $^{72-75}$. In the case of the 3-dim axially symmetric
oblate oscillator with frequency ratio $1:2$ (which corresponds to the case
$m_1 = m_2 =2$, $m_3=1$) the relevant symmetry is related to O(4)
$^{61,76}$. The search for further symmetries, related to specific
frequency ratios, hidden in the general nonlinear algebraic framework given
in this work is an interesting problem.

One of the authors (DB) has been supported by the EU under contract
ERBCHBGCT930467.


\section{References}

\begin{thebibliography}{99}
\bibitem{Dri}  V. G. Drinfeld, in {\it Proceedings of
the International Congress of Mathematicians}, ed. A. M. Gleason
(American Mathematical Society, Providence, RI, 1986), p. 798.

\bibitem{Jim}  M. Jimbo, {\it Lett. Math. Phys.} {\bf 11} (1986) 247.

\bibitem{Abe} E. Abe, {\it Hopf Algebras} (Cambridge University Press,
Cambridge, 1977).

\bibitem{Bie}  L. C.  Biedenharn, {\it J. Phys. A} {\bf 22} (1989) L873.

\bibitem{Mac}  A. J. Macfarlane, {\it J. Phys. A} {\bf 22} (1989) 4581.

\bibitem{Sun}  C. P. Sun and H. C. Fu, {\it J. Phys. A} {\bf 22} (1989)
 L983.

\bibitem{Ari} M.  Arik and D. D. Coon, {\it J. Math. Phys.} {\bf 17} (1976)
 524.

\bibitem{Kur} V. V. Kuryshkin, {\it  Annales de la Fondation Louis de Broglie}
{\bf 5} (1980) 111.

\bibitem{Dask}  C. Daskaloyannis, {\it J. Phys. A} {\bf 24} (1991) L789.

\bibitem{Arik}  M. Arik, E. Demircan, T. Turgut, L. Ekinci, and M. Mungan,
{\it Z. Phys. C} {\bf 55} (1992) 89.

\bibitem{Brz}  T. Brzezi\'nski, I. L. Egusquiza, and A. J. Macfarlane,
{\it Phys. Lett. B}  {\bf 311} (1993) 202.

\bibitem{PLB307}  D.  Bonatsos and C. Daskaloyannis, {\it Phys. Lett. B}
 {\bf 307} (1993) 100.

\bibitem{Mel}  S. Meljanac, M. Milekovic, and S. Pallua, {\it Phys. Lett. B}
{\bf 328} (1994) 55.

\bibitem{Kol}  D. Bonatsos, C. Daskaloyannis, and P. Kolokotronis, {\it J.
Phys. A} {\bf 26} (1993) L871.

\bibitem{DQ1}  C. Delbecq and C. Quesne, {\it J. Phys. A} {\bf 26}
 (1993) L127.

\bibitem{DQ2}  C. Delbecq and C. Quesne, {\it Phys. Lett. B} {\bf 300}
(1993) 227.

\bibitem{DQ3}  C. Delbecq and C. Quesne, {\it Mod. Phys. Lett. A} {\bf 8}
(1993) 961.

\bibitem{LG}  A. Ludu and R. K. Gupta, {\it J. Math. Phys.} {\bf 34}
(1993) 5367.

\bibitem{SunLi} C. P.  Sun and W. Li, {\it Commun. Theor. Phys.}
 {\bf 19} (1993) 191.

\bibitem{Pan}  F. Pan, {\it J. Math. Phys.} {\bf 35} (1994) 5065.

\bibitem{RRS}  P. P. Raychev, R. P. Roussev and Yu. F. Smirnov, {\it J.
Phys. G} {\bf 16} (1990) L137.

\bibitem{PLBVMI} D. Bonatsos, E. N. Argyres, S. B. Drenska, P. P. Raychev,
R. P. Roussev and Yu. F. Smirnov, {\it Phys. Lett. B} {\bf 251} (1990) 477.

\bibitem{JPGSD} D. Bonatsos, S. B. Drenska, P. P. Raychev, R. P. Roussev and
Yu. F. Smirnov, {\it J. Phys. G} {\bf 17} (1991) L67.

\bibitem{CPL175} D. Bonatsos, P. P. Raychev, R. P. Roussev and Yu. F.
Smirnov, {\it Chem. Phys. Lett.} {\bf 175} (1990) 300.

\bibitem{BE2s} D. Bonatsos, A. Faessler, P. P. Raychev, R. P. Roussev and
Yu. F. Smirnov, {\it J. Phys. A} {\bf 25} (1992) 3275.

\bibitem{Minkov} N. Minkov, R. P. Roussev and P. P. Raychev, {\it J. Phys.
G} {\bf 20} (1994) L67.

\bibitem{VIRO} D. Bonatsos, C. Daskaloyannis, A. Faessler, P. P. Raychev
and R. P. Roussev, {\it Phys. Rev. C} {\bf 50} (1994) 497.

\bibitem{Bon} D. Bonatsos, {\it J. Phys. A} {\bf 25} (1992) L101.

\bibitem{BDPLB} D. Bonatsos and C. Daskaloyannis, {\it Phys. Lett. B}
{\bf 278} (1992) 1.

\bibitem{BDF} D. Bonatsos, C. Daskaloyannis and A. Faessler, {\it J. Phys. A}
{\bf 27} (1994) 1299.

\bibitem{BFRRS} D. Bonatsos, A. Faessler, P. P. Raychev, R. P. Roussev
and Yu. F. Smirnov, {\it J. Phys. A} {\bf 25} (1992) L267.

\bibitem{MAP} D. P. Menezes, S. S. Avancini and C. Provid\^encia,
{\it J. Phys. A} {\bf 25} (1992) 6317.

\bibitem{RPA} D. Bonatsos, L. Brito, D. P. Menezes, C. Provid\^encia and
J. da Provid\^encia, {\it J. Phys. A} {\bf 26} (1993) 895, 5185.

\bibitem{HSM} C. Provid\^encia, L. Brito, J. da Provid\^encia, D. Bonatsos
and D. P. Menezes, {\it J. Phys. G} {\bf 20} (1994) 1209.

\bibitem{Mot} B. Mottelson, {\it Nucl. Phys. A} {\bf 522} 1c.

\bibitem{Rae1} W. D. M. Rae, {\it Int. J. Mod. Phys. A} {\bf 3} (1988) 1343.

\bibitem{NoTw} P. J. Nolan and P. J. Twin, {\it Ann. Rev. Nucl. Part. Sci.}
{\bf 38} (1988) 533.

\bibitem{JaKh} R. V. F. Janssens and T. L. Khoo, {\it Ann. Rev. Nucl. Part.
Sci. } {\bf 41} (1991) 321.

\bibitem{RZ} W. D. M. Rae and J. Zhang, {\it Mod. Phys. Lett. A} {\bf 9}
(1994) 599.

\bibitem{ZRM} J. Zhang, W. D. M. Rae and A. C. Merchant, {\it Nucl. Phys. A}
{\bf 575} (1994) 61.

\bibitem{BlBr} D. M. Brink, in {\it Proc. Int. School of Physics, Enrico
Fermi Course XXXVI, Varenna 1966}, ed. C. Bloch (Academic Press, New York,
1966),  p. 247.

\bibitem{MBGL} T. P. Martin, T. Bergmann, H. G\"ohlich and T. Lange,
{\it Z. Phys. D} {\bf 19} (1991) 25.

\bibitem{BL} A. Bulgac and C. Lewenkopf, {\it Phys. Rev. Lett. } {\bf 71}
(1993) 4130.

\bibitem{IBM} F. Iachello and A. Arima, {\it The Interacting Boson Model}
(Cambridge University Press, Cambridge, 1987).

\bibitem{Bonbook} D. Bonatsos, {\it Interacting Boson Models of Nuclear
Structure} (Clarendon, Oxford, 1988).

\bibitem{BK} D. Bonatsos and A. Klein, {\it Phys. Rev. C} {\bf 29} (1984) 1879.

\bibitem{HL} P. Holmberg and P. Lipas, {\it Nucl. Phys. A}  {\bf 117} (1968)
552.

\bibitem{qvibr} R. N. Alvarez, D. Bonatsos and Yu. F. Smirnov, {\it Phys.
Rev. A} {\bf 50} (1994) 1088.

\bibitem{JH}  J. M. Jauch and E. L. Hill, {\it Phys. Rev.} {\bf 57}
 (1940) 641.

\bibitem{Dem}  Yu. N. Demkov, {\it Soviet Phys. JETP} {\bf 17} (1963) 1349.

\bibitem{Con}  G. Contopoulos, {\it Z. Astrophys.} {\bf 49} (1960) 273;
{\it Astrophys. J.} {\bf 138} (1963) 1297.

\bibitem{Cis}  A. Cisneros and H. V. McIntosh, {\it J. Math. Phys.} {\bf 11}
 (1970) 870.

\bibitem{CLP} O.  Casta\~nos R.  and L\'opez-Pe\~na, {\it J. Phys. A}
 {\bf 25} (1992) 6685.

\bibitem{GKM}  A. Ghosh, A. Kundu and P. Mitra, Saha Institute preprint
SINP/TNP/92-5 (1992).

\bibitem{DZ} F.  Duimio and G. Zambotti, {\it Nuovo Cimento} {\bf 43}
(1966) 1203.

\bibitem{Mai}  G. Maiella, {\it Nuovo Cimento} {\bf 52} (1967) 1004.

\bibitem{Ven} I. Vendramin, {\it Nuovo Cimento} {\bf 54} (1968) 190.

\bibitem{MV}  G. Maiella and G. Vilasi, {\it Lettere Nuovo Cimento}
 {\bf 1} (1969) 57.

\bibitem{RD}  G. Rosensteel and J. P. Draayer, {\it J. Phys. A}  {\bf 22}
(1989) 1323.

\bibitem{BCD} D.  Bhaumik, A. Chatterjee and B. Dutta-Roy, {\it  J. Phys. A}
{\bf 27} (1994) 1401.

\bibitem{Naz}  W. Nazarewicz and J. Dobaczewski, {\it Phys. Rev. Lett.}
 {\bf 68} (1992) 154.

\bibitem{Hie}  J. Hietarinta, {\it Phys. Rep.} {\bf 147} (1987) 87.

\bibitem{Evans} N. W. Evans, {\it Phys. Rev. A} {\bf 41} (1990) 5666.

\bibitem{Holt}  C. R. Holt, {\it J. Math. Phys.} {\bf 23} (1982) 1037.

\bibitem{BW}  C. P. Boyer and K. B. Wolf, {\it J. Math. Phys.} {\bf 16}
(1975) 2215.

\bibitem{FL}  A. S. Fokas and P. A. Lagerstrom, {\it J. Math. Anal. Appl.}
 {\bf 74} (1980) 325.

\bibitem{Predeal} D. Bonatsos, C. Daskaloyannis, P. Kolokotronis and D.
Lenis, in {\it Proceedings of the International Summer School on Collective
Motion and Nuclear Dynamics (Predeal 1995)}, ed. A. A. Raduta and D.
Bucurescu (World Scientific, Singapore), in press.

\bibitem{BDK} D.  Bonatsos, C.  Daskaloyannis and K. Kokkotas, {\it Phys. Rev.
A} {\bf 48}  (1993) R3407; {\bf 50} (1994) 3700.

\bibitem{Ell} D. Bonatsos, C. Daskaloyannis, D. Ellinas and A. Faessler,
{\it Phys. Lett. B} {\bf 331} (1994) 150.

\bibitem{Zoup} A. A. Kehagias and G. Zoupanos, {\it Z. Phys. C} {\bf 62}
(1994) 121.

\bibitem{Que} C. Quesne, {\it Phys. Lett. A} {\bf 193} (1994) 245.

\bibitem{Tj1}  T. Tjin, {\it Phys. Lett. B} {\bf 292} (1992) 60.

\bibitem{Tj2}  J. de Boer and T. Tjin, {\it Commun. Math. Phys.}
 {\bf 158} (1993) 485.

\bibitem{Tj3}  T. Tjin, Ph.D. thesis, U. Amsterdam (1993).

\bibitem{Tj4} J. de Boer, F. Harmsze and T. Tjin, SUNY at Stony Brook
preprint ITP-SB-95-07, hep-th 9503161.

\bibitem{Raven} D. G. Ravenhall, R. T. Sharp and W. J. Pardee, {\it Phys.
Rev.} {\bf 164} (1967) 1950.

\bibitem{Nils} S. G. Nilsson, {\it K. Dan. Vidensk. Selsk. Mat. Fys. Medd.}
{\bf 29} (1955) No 16.

\bibitem{Cast} O. Casta\~nos, M. Moshinsky and C. Quesne, in {\it Group
Theory and Special Symmetries in Nuclear Physics (Ann Arbor, 1991)}, ed.
J. P. Draayer and J. W. J\"anecke (World Scientific, Singapore, 1992) p. 80.

\bibitem{Cast2} O. Casta\~nos, M. Moshinsky and C. Quesne, {\it Phys. Lett. B}
{\bf 277} (1992) 238.

\bibitem{Draay} J. P. Draayer, C. Bahri and S. Moszkowski, {\it Rev. Mex.
Fis.} {\bf 38 Supl. 2} (1992) 26.

\bibitem{MosM} M. Moshinsky and A. Del Sol Mesa, {\it Rev. Mex. Fis.}
{\bf 38 Supl. 2} (1992) 146.

\bibitem{Casta} O. Casta\~nos, V. Val\'azquez A., P. O. Hess and J. G. Hirsch,
{\it Phys. Lett.} {\bf 321B} (1994) 303.

\bibitem{DelSol} A. Del Sol Mesa, G. Loyola, M. Moshinsky and V. Vel\'azquez,
{\it J. Phys. A} {\bf 26} (1993) 1147.

\bibitem{Bala} A. B. Balantekin, O. Casta\~nos and M. Moshinsky, {\it Phys.
Lett. B} {\bf 284} (1992) 1.

\bibitem{Smir} Yu. F. Smirnov, V. N. Tolstoy and Yu. I. Kharitonov, {\it
Yad. Fiz. } {\bf 54} (1991) 721 [{\it Sov. J. Nucl. Phys. } {\bf 54}
(1991) 437].

\bibitem{Smir2} Yu. F. Smirnov and Yu. I. Kharitonov, {\it Yad. Fiz. }
{\bf 56} (1993) 263 [{\it Phys. At. Nucl.} {\bf 56} (1993) 1143].

\bibitem{Bohr} A. Bohr and B. R. Mottelson, {\it Nuclear Structure} Vol. II
(Benjamin, Reading, 1975).

\end{thebibliography}
\end{document}